# Stability of Digital Robust Motion Control Systems with Disturbance Observer

Emre Sariyildiz
School of Mechanical, Materials, Mechatronic, and Biomedical Engineering,
Faculty of Engineering and Information Sciences
University of Wollongong
Wollongong, NSW, 2522, Australia
emre@uow.edu.au

*Abstract*— **In this paper, new stability analysis methods are proposed for digital robust motion control systems implemented using a Disturbance Observer (DOb). The proposed methods utilise time domain and traditional frequency domain analysis techniques, and the same stability constraints on the free control parameters of the digital DOb are established through analysis. This paper demonstrates that the observer gain of the digital DOb (aka digital DOb's bandwidth) is constrained by the stability of the observer, in addition to its limitation by noise sensitivity. As the gain of the digital observer is increased, the disturbance estimation accuracy can be enhanced for a limited bandwidth range, and the DOb becomes unstable when the observer gain exceeds the stability constraint derived in the paper. It is also shown that the observer gain's stability constraint changes when disturbances are cancelled through the use of estimated disturbances in an inner-loop. The nominal plant model of the servo system, along with the gain of the digital DOb, must be carefully tuned to ensure good performance and robust stability. The stability constraints identified in this paper can be utilised to effectively adjust the observer gain and nominal plant model of a servo system in DOb-based digital robust motion control, facilitating good performance in practical engineering applications. Results from simulations are presented to verify the novel stability analysis methods.**

*Keywords—Disturbance Observer, Robust Motion Control, and Robust Stability and Performance.*

## I. INTRODUCTION

Motion control systems suffer from various disturbances, e.g., unmodeled dynamics such as backlash and nonlinear frictions in speed reducers, parametric uncertainties such as inertia variations, and external loads when a robot picking and placing different objects [1–5]. To achieve high-performance and robust stability, it is essential to suppress these disturbances in motion control applications [6–8]. Therefore, robustness against disturbances, alongside performance, is a critical requirement for motion control systems in practice [6–10].

To achieve high-performance and robust stability in practice, many robust control methods have been proposed since early 1970s [2–3, 11]. For example, the conventional PID controller can provide good robustness for many practical engineering applications [11, 12]. However, their applications are generally limited to relatively simple engineering systems [9, 11, 12]. To this end, advanced robust control methodologies (e.g., Sliding Mode Control, H∞ control and Internal Model Control) were proposed in the literature [13–19]. Despite significant improvements in this area, complex controller structures and implementation difficulties remain great challenges in many robust motion control methods [9, 12].

Due to its simplicity and low-computational requirements, DOb has been widely adopted by motion control society since its introduction by Ohnishi in 1983 [20]. In this robust control method, disturbances in motion control systems are estimated using a DOb, and the controller's robustness is achieved by compensating for these disturbances with their estimated values [21–23]. To accurately estimate disturbances, DOb's free control parameters, such as the observer gain adjusting the estimation bandwidth and nominal plant model, should be properly tuned [24–26]. These design parameters are typically tuned through trial and error, with consideration primarily given to disturbance estimation accuracy and sensitivity to measurement noise [25]. This oversimplified design approach, however, may lead to poor performance and stability problems in practice [27–29]. To tackle this problem, many researchers have analysed the stability of the robust motion controllers, employing mainly continuous-time analysis methods [28–31]. However, analysis performed in the continuous-time domain cannot fully capture the critical dynamic behaviours of DObs that affect stability when implemented with digital controllers, such as computers. [32–35]. For example, increasing the bandwidth of a DOb can lead to instability in a DOb-based digital robust motion controller, a dynamic behaviour that continuous-time analysis methods fail to address. [34, 35]. Therefore, the stability analysis of DOb-based digital robust motion controllers should be conducted using discrete-time methods. Despite some studies addressing this problem, there remains a deficiency in practical analysis and synthesis methods for digital robust motion controllers with DOb.

To this end, this paper proposes novel discrete-time analysis approaches for the stability problem of digital robust motion control systems based on DObs. It clearly explains how the observer gain and nominal plant parameters change the stability of the digital robust motion controller. The stability constraints on the design parameters of the digital DOb are analytically derived using frequency and time domain analysis methods. The proposed stability constraints serve as highly useful, systematic design tools for achieving high-performance robust motion control systems. Simulation results confirm the effectiveness of the proposed analysis and synthesis methods.

The remainder of the paper is structured as follows. Section II introduces a general dynamic model for motion control systems in state space, covering both the continuous- and discrete- time domain implementations. Section III explains the conventional DOb synthesis in the discrete-time domain. Section IV proposes two stability analysis methods, utilising frequency and time domain approaches. Section V presents simulation results, while Section VI concludes the paper.

## II. DYNAMIC MODEL OF A SERVO SYSTEM

*a) Dynamic Model in the Continuous-time Domain:*

Equation (1) provides the state space model of a servo system in continuous-time.

$$\dot{\mathbf{x}}(t) = \mathbf{A}_\mathbf{C}\mathbf{x}(t) + \mathbf{B}_\mathbf{C}u(t) - \mathbf{D}_\mathbf{C}\tau_d(t) \quad (1)$$

where $\mathbf{A}_\mathbf{C} = \begin{bmatrix} 0 & 1 \\ 0 & b_m/J_m \end{bmatrix}$ represents the state matrix of the dynamic model in which $J_m$ and $b_m$ are the inertia and viscous friction coefficient of the servo system, respectively; $\mathbf{B}_\mathbf{C} = [0 \ 1/J_m]$ represents the control input vector; $\mathbf{D}_\mathbf{C} = [0 \ 1/J_m]$ represents the disturbance input vector; $\mathbf{x}(t) = [q(t) \ \dot{q}(t)]^T$ represents the state vector of the servo system in which $q(t)$ and $\dot{q}(t)$ are the position and velocity states, respectively; $\dot{\mathbf{x}}(t)$ represents the derivative of the state vector; $u(t)$ represents the control input generated by the motion controller; $\tau_d(t)$ represents external disturbances such as load and unmodeled dynamics, e.g., nonlinear friction and hysteresis in speed reducers; and $t$ represents time [19, 28].

When the exact inertia and viscous friction parameters cannot be identified, the following nominal dynamic model can be used for servo systems.

$$\dot{\mathbf{x}}(t) = \mathbf{A}_\mathbf{Cn}\mathbf{x}(t) + \mathbf{B}_\mathbf{Cn}u(t) - \mathbf{D}_\mathbf{Cn}\tau_{dn}(t) \quad (2)$$

where $\mathbf{A}_\mathbf{Cn} = \begin{bmatrix} 0 & 1 \\ 0 & b_{mn}/J_{mn} \end{bmatrix}$, represents the nominal state matrix in which $J_{mn}$ and $b_{mn}$ are the nominal inertia and viscous friction coefficient, respectively; $\mathbf{B}_\mathbf{Cn} = [0 \ 1/J_{mn}]^T$ represents the nominal control input vector; $\mathbf{D}_\mathbf{Cn} = [0 \ 1/J_{mn}]^T$ represents the nominal disturbance input vector; and $\tau_{dn}(t)$ represents the nominal disturbance variable, including external disturbances, unmodeled dynamics and parametric uncertainties as follows:

$$\tau_{dn}(t) = \frac{1}{\mathbf{D}_\mathbf{Cn}^T\mathbf{D}_\mathbf{Cn}}\left((\mathbf{A}_\mathbf{Cn} - \mathbf{A}_\mathbf{C})\mathbf{x}(t) + (\mathbf{B}_\mathbf{Cn} - \mathbf{B}_\mathbf{C})u(t) + \mathbf{D}_\mathbf{C}\tau_d(t)\right) \quad (3)$$

*b) Dynamic Model in the Discrete-time Domain:*

The solution of the differential equation given in Eq. (1) is as follows:

$$\mathbf{x}(t) = e^{\mathbf{A}_\mathbf{C}(t-t_0)}\mathbf{x}(t_0) + \int_{t_0}^{t} e^{\mathbf{A}_\mathbf{C}(t-\lambda)}\left(\mathbf{B}_\mathbf{C}u(\lambda) - \mathbf{D}_\mathbf{C}\tau_d(\lambda)\right)d\lambda \quad (4)$$

where $e^\bullet$ represents the exponential of a matrix $\bullet$, and $t_0$ represents the initial time.

To obtain the discrete model of the servo system, let us use $t_0 = kT_s$ and $t = (k+1)T_s$ where $T_s$ is the sampling period and $k$ is the discrete-time index.

$$\mathbf{x}((k+1)T_s) = e^{\mathbf{A}_\mathbf{C}T_s}\mathbf{x}(kT_s) + \int_{kT_s}^{(k+1)T_s} e^{\mathbf{A}_\mathbf{C}((k+1)T_s-\lambda)}\left(\mathbf{B}_\mathbf{C}u(\lambda) - \mathbf{D}_\mathbf{C}\tau_d(\lambda)\right)d\lambda \quad (5)$$

Let us change the integral variable using $\tau = (k+1)T_s - \lambda$. When the new integral variable is substituted into Eq. (5), the discrete model of the servo system is derived as follows:

$$\mathbf{x}((k+1)T_s) = \mathbf{A}_\mathbf{D}\mathbf{x}(kT_s) + \mathbf{B}_\mathbf{D}u(kT_s) - \mathbf{\Pi}_\mathbf{D}(kT_s) \quad (6)$$

where $\mathbf{A}_\mathbf{D} = e^{\mathbf{A}_\mathbf{C}T_s}$ represents the discrete state matrix of the servo system, $\mathbf{B}_\mathbf{D} = \int_0^{T_s} e^{\mathbf{A}_\mathbf{C}\tau}\mathbf{B}_\mathbf{C}d\tau$ represents the discrete control input vector, and $\mathbf{\Pi}_\mathbf{D}(kT_s) = \int_0^{T_s} e^{\mathbf{A}_\mathbf{C}\tau}\mathbf{D}_\mathbf{C}\tau_d((k+1)T_s - \tau)d\tau$ represents the discrete disturbance vector.

Equation (7) similarly represents the discrete model using nominal plant parameters instead of exact ones.

$$\mathbf{x}((k+1)T_s) = \mathbf{A}_\mathbf{Dn}\mathbf{x}(kT_s) + \mathbf{B}_\mathbf{Dn}u(kT_s) - \mathbf{\Pi}_\mathbf{Dn}(kT_s) \quad (7)$$

where $\mathbf{A}_\mathbf{Dn} = e^{\mathbf{A}_\mathbf{Cn}T_s}$ is the discrete nominal state matrix of the servo system, $\mathbf{B}_\mathbf{Dn} = \int_0^{T_s} e^{\mathbf{A}_\mathbf{Cn}\tau}\mathbf{B}_\mathbf{Cn}d\tau$ is the discrete nominal control input vector, and $\mathbf{\Pi}_\mathbf{Dn}(kT_s) = \int_0^{T_s} e^{\mathbf{A}_\mathbf{Cn}\tau}\mathbf{D}_\mathbf{Cn}\tau_{dn}((k+1)T_s - \tau)d\tau$ is the discrete nominal disturbance vector [36–39].

## III. DISTURBANCE OBSERVER-BASED ROBUST CONTROL

*a) Disturbance Observer*

To design a DOb, a dynamic model should be introduced for the nominal disturbance vector in Eq. (7) [8]. For example, the conventional DOb can be designed by assuming that the nominal disturbance variable does not change within the sampling period. Substituting this assumption into Eq. (7) yields

$$\mathbf{x}((k+1)T_s) = \mathbf{A}_\mathbf{Dn}\mathbf{x}(kT_s) + \mathbf{B}_\mathbf{Dn}u(kT_s) - \mathbf{D}_\mathbf{Dn}\tau_{dn}(kT_s) \quad (8)$$

where $\mathbf{D}_\mathbf{Dn} = \int_0^{T_s} e^{\mathbf{A}_\mathbf{Cn}\tau}\mathbf{D}_\mathbf{Cn}d\tau$ is the discrete nominal disturbance input vector, and $\tau_{dn}(kT_s)$ represents the nominal disturbance variable at $kT_s$ seconds. The other parameters are the same as those defined earlier.

Let us employ the following auxiliary variable to design the conventional DOb.

$$z(kT_s) = \tau_{dn}(kT_s) + \mathbf{L}^T\mathbf{x}(kT_s) \quad (9)$$

where $\mathbf{L} \in R^2$ is the observer gain vector. The stability and performance of disturbance estimation are adjusted by tuning the observer gain vector as shown in Section IV.

The dynamics of the auxiliary variable is derived by substituting Eq. (8) into Eq. (9) as follows:

$$\begin{aligned}
z((k+1)T_s) &= \tau_{dn}((k+1)T_s) + \mathbf{L}^T\mathbf{x}((k+1)T_s) \\
&= \mathbf{L}^T\left(\mathbf{A}_\mathbf{Dn}\mathbf{x}(kT_s) + \mathbf{B}_\mathbf{Dn}u(kT_s) - \mathbf{D}_\mathbf{Dn}\tau_{dn}(kT_s)\right) + \Delta\tau_{dn}(kT_s) \\
&= (1 - \mathbf{L}^T\mathbf{D}_\mathbf{Dn})z(kT_s) + \mathbf{L}^T\left(\mathbf{A}_\mathbf{Dn} + \mathbf{D}_\mathbf{Dn}\mathbf{L}^T - \mathbf{I}_2\right)\mathbf{x}(kT_s) + \\
&\quad \mathbf{L}^T\mathbf{B}_\mathbf{Dn}u(kT_s) + \Delta\tau_{dn}(kT_s)
\end{aligned} \quad (10)$$

where $\Delta\tau_{dn}(kT_s) = \tau_{dn}((k+1)T_s) - \tau_{dn}(kT_s)$ is the difference between the nominal disturbance variable at $kT_s$ and $(k+1)T_s$ seconds, and $\mathbf{I_2}$ is a 2x2 identity matrix.

The dynamics of the auxiliary variable observer is obtained by neglecting $\Delta\tau_{dn}(kT_s)$ in Eq. (10) as follows:

$$\hat{z}((k+1)T_s) = (1-\mathbf{L}^T\mathbf{D_{Dn}})\hat{z}(kT_s) + \mathbf{L}^T(\mathbf{A_{Dn}} + \mathbf{D_{Dn}}\mathbf{L}^T - \mathbf{I_2})\mathbf{x}(kT_s) + \mathbf{L}^T\mathbf{B_{Dn}}u(kT_s) \quad (11)$$

where $\hat{z}(kT_s)$ represents the estimated auxiliary variable at $kT_s$ seconds.

When the observer gain vector is properly tuned, the estimated auxiliary variable converges to the auxiliary variable given in Eq. (9), following the governing dynamics of the observer system. Hence, the nominal disturbance variable can be estimated using Eq. (12).

$$\hat{\tau}_{dn}(kT_s) = \hat{z}(kT_s) - \mathbf{L}^T\mathbf{x}(kT_s) \quad (12)$$

where $\hat{\tau}_{dn}(kT_s)$ represents the estimated nominal disturbance variable at $kT_s$ seconds.

*b) Robust Motion Controller*

A DOb-based robust motion controller is synthesised using two feedback-control loops, namely inner- and outer- feedback control loops. While the robustness of servo systems is achieved by feeding back the estimated disturbances in the inner-loop, the outer-loop controller is responsible for regulating the overall system performance to meet desired dynamic specifications [6].

Let us first consider the inner-loop. When the estimated disturbance is fed back in the inner-loop, the dynamic model of the robust motion controller is derived by combining Eqs. (8) and (11) as follows:

$$\mathbf{x_a}((k+1)T_s) = \mathbf{A_{Di}}\mathbf{x_a}(kT_s) + \mathbf{B_{Di}}u_p(kT_s) - \mathbf{\Pi_{Di}}(kT_s) \quad (13)$$

where $\mathbf{x_a}(kT_s) = [\mathbf{x}(kT_s) \quad z(kT_s)]^T$ represents the augmented state vector, comprising the servo states and auxiliary variable, $\mathbf{A_{Di}} = \begin{bmatrix} \mathbf{A_D} - \mathbf{B_D}\mathbf{L}^T & \mathbf{B_D} \\ \mathbf{L}^T(\mathbf{A_{Dn}} - \mathbf{I_2}) & 1 \end{bmatrix}$ represents the augmented state matrix in the inner-loop, $\mathbf{B_{Di}} = [\mathbf{B_D}^T \quad \mathbf{L}^T\mathbf{B_{Dn}}]^T$ represents the augmented control input vector in the inner-loop, $\mathbf{\Pi_{Di}}(kT_s) = [\mathbf{\Pi_D}^T(kT_s) \quad 0]^T$ represents the augmented disturbance vector in the inner-loop, and $u_p(k)$ is the performance control input.

Let us now consider the outer-feedback control loop and employ the following state feedback controller to adjust the performance of the motion control system.

$$u_p(kT_s) = q^{ref}(kT_s) - \mathbf{K}^T\mathbf{x_a}(kT_s) \quad (14)$$

where $q^{ref}(kT_s)$ represents the reference position of the servo system, and $\mathbf{K} = [K_p \quad K_v \quad 0]^T \in R^3$ represents the control gain vector in which $K_p$ and $K_v \in R$ are the position and velocity gains, respectively.

Substituting Eq. (14) into Eq. (13) yields

$$\mathbf{x_a}((k+1)T_s) = \mathbf{A_{Do}}\mathbf{x_a}(kT_s) + \mathbf{B_{Do}}q^{ref}(kT_s) - \mathbf{\Pi_{Do}}(kT_s) \quad (15)$$

where $\mathbf{A_{Do}} = \begin{bmatrix} \mathbf{A_D} - \mathbf{B_D}(\mathbf{L}^T + \mathbf{K}^T) & \mathbf{B_D} \\ \mathbf{L}^T(\mathbf{A_{Dn}} - \mathbf{I_2}) & 1 \end{bmatrix}$ represents the augmented state matrix in the outer-loop, $\mathbf{B_{Do}} = [\mathbf{B_D}^T \quad \mathbf{L}^T\mathbf{B_{Dn}}]^T$ represents the augmented control input vector in the outer-loop, and $\mathbf{\Pi_{Do}}(kT_s) = [\mathbf{\Pi_D}^T(kT_s) \quad 0]^T$ represents the augmented disturbance vector in the outer-loop.

IV. STABILITY ANALYSIS

By subtracting Eq. (11) from Eq. (10), the estimation error dynamics can be obtained as follows:

$$e_z((k+1)T_s) = (1-\mathbf{L}^T\mathbf{D_{Dn}})e_z(kT_s) + \Delta\tau_{dn}(kT_s) \quad (16)$$

where $e_z(kT_s) = z(kT_s) - \hat{z}(kT_s)$ represents the auxiliary variable estimation error at $kT_s$ seconds.

Equation (16) shows that the estimation error converges to a bounded value when the absolute value of $1-\mathbf{L}^T\mathbf{D_{Dn}}$ is smaller than one. The upper bound of the estimation error is determined by $\Delta\tau_{dn}(kT_s)$, and the estimation error converges to zero when $\Delta\tau_{dn}(kT_s)$ is null, i.e., $\tau_{dn}((k+1)T_s) = \tau_{dn}(kT_s)$. Therefore, a stability constraint on the free control parameters of observer gain vector is obtained as follows:

$$0 < g_D < 2/|\mathbf{D_{Dn}}|_1 \quad (17)$$

where $g_D$ is a free control parameter in the observer gain vector, i.e., $\mathbf{L} = g_D[1 \quad 1]^T$, and $|\bullet|_1$ is the L1 norm of vector $\bullet$.

Equation (17) provides a basic insight into the stability of DOb-based digital robust motion control systems. However, it falls short in explaining certain dynamic responses because the stability constraints of the robust motion controller change when the estimated disturbance is fed back into the inner-loop. Let us now consider the stability of inner- and outer- loops of the digital robust motion controller.

*a) Frequency Domain Analysis:*

The state space model of the robust motion controller can be described in the inner-loop as follows:

$$\mathbf{x_a}((k+1)T_s) = \mathbf{A_{Di}}\mathbf{x_a}(kT_s) + \mathbf{B_{Di}}u_p(kT_s) - \mathbf{\Pi_{Di}}(kT_s)$$
$$y_i = \mathbf{C_{Di}}^T\mathbf{x_a}(kT_s) \quad (18)$$

where $y_i \in R$ represents the output, $\mathbf{C_{Di}} \in R^3$ represents the output vector, and the other parameters are described in Eq. (13).

When the viscous friction term is neglected, the transfer function between the control input and the velocity state of the servo system can be derived using Eq. (19).

$$\frac{\dot{q}(z)}{u_p(z)} = \mathbf{C}_{\mathbf{Di}}^T (z\mathbf{I}_3 - \mathbf{A}_{\mathbf{Di}})^{-1} \mathbf{B}_{\mathbf{Di}} \quad (19)$$
$$= \frac{T_s}{J_m} \frac{(z-1+g_D)}{(z-1)(z-1+\alpha g_D)}$$

where $\mathbf{C}_{\mathbf{Di}} = [0 \ 1 \ 0]^T$ and $\alpha = J_{mn}/J_m$.

Equation (19) shows that the inner-loop transfer function has a pole at $1-\alpha g_D$. Therefore, the nominal inertia, as well as the observer gain, must be carefully tuned to achieve stability. From Eq. (19), the constraint on the design parameters of the digital DOb is derived as follows:

$$0 < \alpha g_D < 2 \quad (20)$$

In the outer-loop, the state space model of the robust motion controller can be similarly described as follows:

$$\mathbf{x}_\mathbf{a}((k+1)T_s) = \mathbf{A}_{\mathbf{Do}}\mathbf{x}_\mathbf{a}(kT_s) + \mathbf{B}_{\mathbf{Do}}u_p(kT_s) - \mathbf{\Pi}_{\mathbf{Do}}(kT_s) \quad (21)$$
$$y_o = \mathbf{C}_{\mathbf{Do}}^T \mathbf{x}_\mathbf{a}(kT_s)$$

where $y_o \in R$ represents the output, $\mathbf{C}_{\mathbf{Do}} \in R^3$ represents the output vector, and the other parameters are described in Eq. (15).

The transfer function between the control input and the velocity state of the servo system can be derived using Eq. (22).

$$\frac{q(z)}{q^{ref}(z)} = \mathbf{C}_{\mathbf{Do}}^T (z\mathbf{I}_3 - \mathbf{A}_{\mathbf{Do}})^{-1} \mathbf{B}_{\mathbf{Do}} = \frac{L_o(z)}{1+L_o(z)}$$
$$L_o(z) = \frac{T_s}{2J_m} \frac{z-(1-g_D)}{z-(1-\alpha g_D)} \frac{2K_d(z-1)+K_pT_s(z+1)}{(z-1)^2} \quad (22)$$

where $\mathbf{C}_{\mathbf{Do}} = [K_p \ K_d \ 0]^T$.

Equation (22) shows that not only the design parameters of the digital DOb but also the gains of the performance controller in the outer-loop affects the stability of the robust motion controller. Therefore, the observer design parameters $\alpha$ and $g_D$ and the performance control gains $K_p$ and $K_d$ should be tuned by considering the stability of the transfer function given in Eq. (22). There is a phase lead/lag compensator in the open-loop transfer function of the robust motion controller. The phase-lead effect can be increased by using the higher values of $\alpha$ in the design of the DOb. However, the phase margin of the robust motion controller cannot be freely adjusted as the open-loop transfer function becomes unstable when $|1-\alpha g_D| > 1$.

*b) State-Space Analysis:*

Let us first consider the inner-loop. When the viscous friction term is neglected, the eigenvalues of the state space model given in Eq. (18) are derived as 1, 1 and $1-\alpha g_D$. Therefore we can obtain the same design constraint given in Eq. (20) using the state space analysis.

Let us now consider the outer-loop of the robust motion controller. The inner-loop state-space model of the robust controller can be rewritten using the Jordan canonical form as follows:

$$\tilde{\mathbf{x}}_\mathbf{a}((k+1)T_s) = \tilde{\mathbf{A}}_{\mathbf{Di}}\tilde{\mathbf{x}}_\mathbf{a}(kT_s) + \tilde{\mathbf{B}}_{\mathbf{Di}}u_p(kT_s) - \tilde{\mathbf{\Pi}}_{\mathbf{Di}}(kT_s) \quad (23)$$
$$y_i = \tilde{\mathbf{C}}_{\mathbf{Di}}^T \tilde{\mathbf{x}}_\mathbf{a}(kT_s)$$

where $\tilde{\mathbf{A}}_{\mathbf{Di}} = \mathbf{M}\mathbf{A}_{\mathbf{Di}}\mathbf{M}^{-1} = \begin{bmatrix} 1 & 1 & 0 \\ 0 & 1 & 0 \\ 0 & 0 & 1-\alpha g_D \end{bmatrix}$ is the state matrix in the Jordan canonical form, $\tilde{\mathbf{x}}_\mathbf{a}(kT_s) = \mathbf{M}\mathbf{x}_\mathbf{a}(kT_s)$ is the state vector, $\tilde{\mathbf{B}}_{\mathbf{Di}} = \mathbf{M}\mathbf{B}_{\mathbf{Di}}$ is the control input vector, $\tilde{\mathbf{\Pi}}_{\mathbf{Di}} = \mathbf{M}\mathbf{\Pi}_{\mathbf{Di}}$ is the disturbance vector and $\tilde{\mathbf{C}}_{\mathbf{Di}}^T = \mathbf{C}_{\mathbf{Di}}^T \mathbf{M}^{-1}$ is the output vector. The transformation matrix $\mathbf{M}$ is given in Eq. (24).

$$\mathbf{M} = \begin{bmatrix} \frac{T_s(2+T_s)}{2g_D J_{mn}} & -\frac{2+T_s}{2g_D J_{mn}} & -\frac{(-2J_m + g_D J_{mn})T_s(2+T_s)}{4g_D J_m J_{mn}} \\ 0 & \frac{2+T_s}{2g_D J_{mn}} & -\frac{2+T_s}{2J_m} \\ 1 & 0 & 1 \end{bmatrix} \quad (24)$$

When $u_p(k) = q^{ref}(kT_s) - \mathbf{K}^T \tilde{\mathbf{x}}_\mathbf{a}(kT_s)$ is substituted into Eq. (23), the outer-loop state space model is obtained as follows:

$$\tilde{\mathbf{x}}_\mathbf{a}((k+1)T_s) = \tilde{\mathbf{A}}_{\mathbf{Do}}\tilde{\mathbf{x}}_\mathbf{a}(kT_s) + \tilde{\mathbf{B}}_{\mathbf{Di}}q^{ref}(kT_s) - \tilde{\mathbf{\Pi}}_{\mathbf{Di}}(kT_s) \quad (25)$$
$$y_o = \tilde{\mathbf{C}}_{\mathbf{Di}}^T \tilde{\mathbf{x}}_\mathbf{a}(kT_s)$$

where $\tilde{\mathbf{A}}_{\mathbf{Do}} = \begin{bmatrix} 1-K_p\left(g_D + \frac{2(J_{mn}-J_m)T_s}{J_{mn}(2+T_s)}\right) & 1-K_d\left(g_D + \frac{2(J_{mn}-J_m)T_s}{J_{mn}(2+T_s)}\right) & 0 \\ \frac{-2g_D K_p T_s}{2+T_s} & 1-\frac{2g_D K_d T_s}{2+T_s} & 0 \\ \frac{2(J_{mn}-J_m)K_p T_s}{J_{mn}(2+T_s)} & \frac{2(J_{mn}-J_m)K_d T_s}{J_{mn}(2+T_s)} & 1-\alpha g_D \end{bmatrix}$,

and the other parameters are given in Eq. (23).

In the outer-loop state matrix, the off diagonal terms $\tilde{\mathbf{A}}_{\mathbf{Do}}(1,3)$ and $\tilde{\mathbf{A}}_{\mathbf{Do}}(2,3)$ are zero. Therefore, the characteristic polynomial of the outer-loop state space model can be obtained using Eq. (26).

$$|\lambda\mathbf{I}_3 - \tilde{\mathbf{A}}_{\mathbf{Do}}| = |\lambda - (1-\alpha g_D)| \times$$
$$\left|\lambda\mathbf{I}_2 - \begin{bmatrix} 1-K_p\left(g_D + \frac{2(J_{mn}-J_m)T_s}{J_{mn}(2+T_s)}\right) & 1-K_d\left(g_D + \frac{2(J_{mn}-J_m)T_s}{J_{mn}(2+T_s)}\right) \\ \frac{-2g_D K_p T_s}{2+T_s} & 1-\frac{2g_D K_d T_s}{2+T_s} \end{bmatrix}\right| \quad (26)$$

where $\mathbf{I}_\bullet$ is a $\bullet \times \bullet$ identity matrix.

Equation (26) shows that the inner and outer-loop eigenvalues can be adjusted separately. While an eigenvalue can be directly set at $1-\alpha g_D$ (i.e., the eigenvalue of the inner-loop feedback controller), the other two eigenvalues are adjusted by tuning the performance control gains $K_p$ and $K_d$.

## V. SIMULATIONS

This section validates the novel discrete stability analysis methods through position control simulations. The following model and controller parameters are employed when the simulations are performed in MATLAB/SIMULINK: $J_m = 0.1$ kg, $T_s = 1$ms, $K_p = 500$ and $K_d = 25$. Different bandwidth and nominal inertia values are used in the simulations.

Figure 1 demonstrates the root-loci of the digital robust position controller when different design parameters are employed in the DOb synthesis. As shown in Fig. 1a and Fig. 1b, the robust motion controller becomes unstable as the observer gain (i.e., the bandwidth of the DOb) is increased. This dynamic behaviour is expected, because Eq. (16) shows that the observer gain cannot be freely tuned due the stability constraint given in Eq. (17). Figures 1a and 1b also show that the nominal inertia has a significant impact on the stability of the robust motion controller. As shown in Eq. (22), a phase lag compensator is synthesised in the inner-loop when the nominal inertia is tuned so that $\alpha<1$. This negatively affects stability, and the closed poles of the robust motion controller tend to move outside the unit circle for low observer gain values. To improve the stability of the robust motion controller, a phase lead compensator should be implemented by tuning $\alpha>1$ in the DOb synthesis. To increase the phase-lead effect, higher values of nominal inertia can be used in the design of the DOb. However, this design parameter cannot be freely increased due to the stability constraint given in Eqs. (19), (20) and (26). As shown in Fig. 1c, the closed-loop poles of the robust motion controller tend to move outside the unit circle, hence degraded stability in robust motion control. To achieve good performance and robust stability, the nominal inertia and observer gain should be tuned using the proposed stability constraints.

Figure 2 illustrates the regulation and trajectory tracking control of the conventional PID and DOb-based robust motion controllers when the external disturbance illustrated in Fig. 2b is applied to a servo system. Simulation results show that the DOb-based robust motion controller outperforms the conventional PID controller. However, the design parameters of the robust motion controller should be properly tuned to achieve good performance and robust stability. As shown in the figure, high-performance motion control tasks can be performed using the proposed stability constraints in the design of the DOb.

## VI. CONCLUSION

In this paper, novel stability analysis methods have been presented for digital robust motion control systems implemented using a DOb within the discrete-time domain. Compared to the conventional analysis methods conducted in the continuous-time domain, the proposed discrete stability analysis shows that the observer gain (i.e., disturbance estimation bandwidth) has an upper bound due to the stability constraints given in Eqs. (17) and (20). These stability constraints are more realistic than the

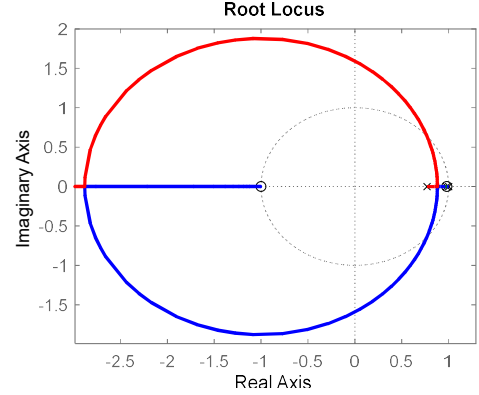

a) Root-loci with respect to $g_D$ when $\alpha<1$.

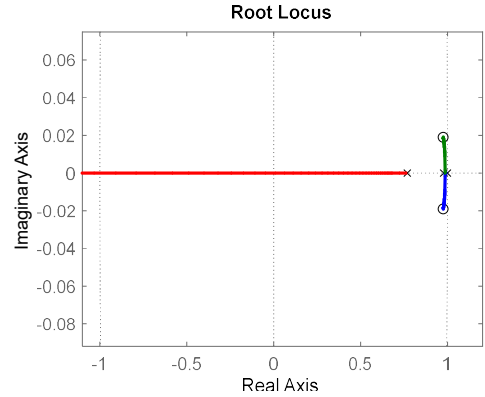

b) Root-loci with respect to $g_D$ when $\alpha>1$.

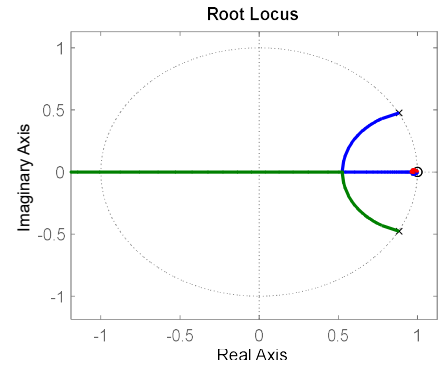

c) Root-loci with respect to $\alpha$.

Fig. 1: Root-loci of the robust motion controller.

ones derived using the conventional stability analysis methods, because they enable us to explain certain dynamic behaviours of the digital robust motion controllers in practice. In addition to the bandwidth of the DOb, the stability and performance of the robust motion controller can be adjusted by tuning the nominal inertia term in the DOb synthesis. As the nominal inertia is increased, the phase-lead effect improves the stability of the robust motion controller. However, the nominal inertia cannot be freely increased as well due to the stability constraint derived in Eqs. (20) and (26). The simulation results substantiate the critical importance of design constraints in ensuring system stability and performance.

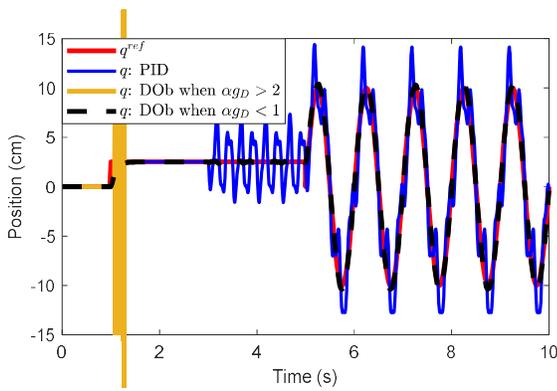

a) Regulation and trajectory tracking control.

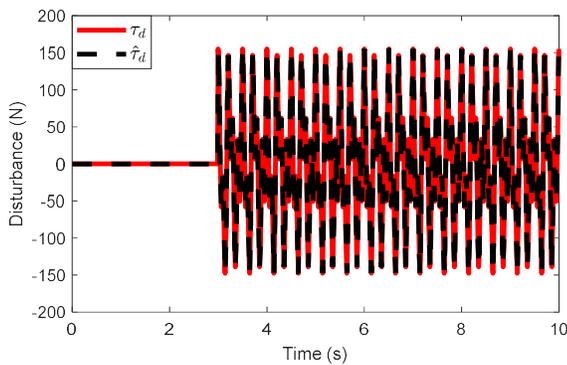

b) Disturbance force and its estimation when $\alpha g_D < 1$.

Fig. 2: Robust position control.